\documentclass[aps, preprint,onecolumn,netbib]{revtex4}
\usepackage{graphics}      
\usepackage{graphicx}      
\usepackage{longtable}     
\usepackage{url}           
\usepackage{bm,color}            
\usepackage{amssymb, amsmath, enumerate, theorem, epsfig,subfigure,setspace}

\begin{document}
\title{Matter wave  coupling of spatially separated and unequally pumped polariton condensates}

\author{Kirill Kalinin$^{1,2}$, Pavlos G. Lagoudakis$^{1,3}$ and Natalia G. Berloff${}^{1,2}$ }
\email[correspondence address: ]{N.G.Berloff@damtp.cam.ac.uk}
\affiliation{$^1$Skolkovo Institute of Science and Technology Novaya St., 100, Skolkovo 143025, Russian Federation}
\affiliation{$^2$Department of Applied Mathematics and Theoretical Physics, University of Cambridge, Cambridge CB3 0WA, United Kingdom }
\affiliation{$^3$Department of Physics and  Astronomy, University of Southampton, Southampton, SO17 1BJ, United Kingdom}

\date{\today}

\begin{abstract}{Spatial quantum coherence between two separated driven-dissipative polariton condensates created non-resonantly and with a different occupation is studied.  
We identify the regions where the condensates remain coherent with the phase difference continuously changing with the pumping imbalance  and the regions where each  condensate acquires its own chemical potential with phase differences exhibiting time-dependent oscillations.  
We show that in the mutual coherence limit the coupling consists of  two competing contributions: a symmetric Heisenberg exchange and  the Dzyloshinskii-Moriya asymmetric interactions that enable a continuous tuning of the phase relation across the dyad and derive analytic expressions for these types of interactions. The introduction of non-equal pumping increases the complexity of  the type of the problems that can be solved by polariton condensates arranged in a graph configuration. If equally pumped polaritons condensates arrange their phases to solve the constrained quadratic minimisation problem with a real symmetric matrix, the non-equally pumped condensates solve that problem for a general Hermitian matrix. }
\end{abstract}

\maketitle

\section{Introduction}
One of the hallmarks of the early discoveries of the Bose-Einstein condensation in a weakly interacting gas of alkali-metal atoms  was the observation of interference fringes in two overlapping condensates  that remarkably demonstrated  the existence of macroscopic spatial quantum coherence \citep{??andrews97}. The superfluidity of these condensates, however, can only be verified by some characteristic signatures  such as the presence of persistent currents. Therefore, many  early theoretical and experimental efforts in ultracold Bose-Einstein condensates (BECs) were focused on the creation of a Josephson junction between two condensates \citep{hall98, dalfovo96} and the studies of the coherent atomic oscillations between two weakly coupled Bose-Einstein condensates \citep{raghavan99}. Such matter-wave interference experiments are important for the understanding matter at its most basic level and  were proposed to be used in  the development of high-precision interferometric sensors. An integrated interferometer based on a coherent matter-wave beam splitter has already been  constructed on an atom chip \cite{schumm05}.

Driven-dissipative condensate systems, such as exciton-polariton (or just polariton) condensates,  are capable of introducing even richer matter wave interference physics than equilibrium systems due to the typical presence of persistent currents even in the steady state configurations. 
Polaritons are mixed light-matter quasiparticles that form  due to the strong coupling of photons in a microcavity and excitons in a semiconductor quantum well \cite{kasprzak}. They are composite bosons that at low densities  can form a
condensed (coherent) state above a critical density \cite{kasprzak, kavokin, keelingberloffreview}. Polariton condensates  are intrinsically nonequilibrium systems with the steady states set by the balance between pumping and losses due to the short lifetime of polariton as photons leak through the confining mirrors. The first theoretical prediction that the non-resonantly pumped polariton condensates separated by the distances larger than the spatial extent of the pumping profile may phase lock \cite{keelingberloffLattices} quickly followed by the experimental demonstration of such locking for three condensates created at the corners of a equilateral triangle  and four condensates arranged at the corners of a square \cite{tosi13} raising an issue about the nature of the phase locking \cite{njp14}. The Josephson oscillations of two coupled trapped spinor polariton condensates were also experimentally achieved \cite{ohadi17} studying  transport-related effects like Bloch oscillations. It was shown that the phase coupling of trapped condensates depends on  the interplay between the Josephson coupling strength and the internal linear polarization splitting of each condensate. However,  in the absence of a potential barrier separating remote polariton condensates (as in trapped condensates), and the concomitant Josephson coupling, interactions between two equally pumped condensates lead to a symmetric Heisenberg exchange type of coupling that imposes either in-phase or anti-phase configuration across a polariton dyad \cite{OGFR}. It was shown that  the phase coupling  depends both on the separation distance and outflow velocity of polaritons from the reservoirs of hot excitons created by incoherent pumping  \cite{OGFR}. In these studies only equally pumped condensates were addressed.

Our interest in the coupling of driven-dissipative condensates is not limited by the goal of the understanding the basic physics of the matter wave interference. Polariton graphs or polariton lattices were recently proposed and realised as  a new physical platform to be used as an analogue Hamiltonian optimizer \citep{BerloffNatMat2017, BerloffNJP2017}, where individual polariton condensates are imprinted into vertices of an arbitrary two-dimensional graph by spatial modulation of the pumping laser.
A large variety of  real life optimization problems can be mapped into certain universal classical spin models such as an Ising, $XY$ or Heisenberg models, in such a way that the variables are mapped into ``spins", their interactions are represented by  ``couplings," and the cost function is represented by a ``Hamiltonian" \citep{cubittScience16}. Various physical platforms have been proposed to simulate universal classical spin models including superconducting qubits \citep{SupercondQubitsNature2011}, optical lattices \citep{OpticalLatticesScience2011}, and photon laser networks \citep{CoupledLasersPRL2013} among others (for the review of these and other systems also see \citep{quantumsimulatorsrev}).  Based on well-established semiconductor and optical control technologies, polariton graph optimizers  benefit from flexible tunability that allows for a variety of coupling strengths between the vertices to be realised by simply adjusting the characteristics of the pump. In the case of optically imprinted polariton lattices with freely propagating polariton condensates, we recently demonstrated that the phase-configuration acquired in a polariton dyad or triad is chosen so as to maximise polariton occupancy \citep{OGFR}, while by expanding to square, and rhombic lattices as well as to arbitrary polariton graphs we simulated minimization of an XY Hamiltonian through bosonic stimulation \citep{BerloffNatMat2017}. The bottom-up approach of bosonic stimulation is achieved in polariton simulators by gradually increasing the excitation density to condensation threshold. This mechanism has a potential  advantage over classical or quantum annealing techniques, where the global ground state is reached through time-dependent transitions over metastable excited states with an exponential growth of the cost of the search with the size of the system \citep{QuantAnnealing1,QuantAnnealing2,QuantAnnealing3,QuantAnnealing4}.

By controlling the separation distance and the in-plane wave-vector we acquire several degrees of freedom in the tunability of inter-site interactions, whereas the continuous coupling of polaritons to free photons allows for in-situ read out of all the characteristics of the polariton condensates such as energy, momentum, and most critically their relative  phase. 
Such flexibility gives the necessary tools to further realize  nontrivial matter states not possible or difficult to observe in other systems. We have already shown the potential of polariton graphs  for creating  discrete giant vortices \citep{GiantVortex} and frustrated states \citep{ExoticStates1D}, controllable next nearest neighbour interactions \citep{ExoticStates1D},  dynamic phase transitions \cite{OGFR}, and realising the spectral gaps \cite{SpectralGap}.

In this paper, we show that  varying the relative population of two polariton condensates adds a new degree of freedom that would allow for an additional control on the coupling between condensates. We analytically derive the closed form expression for the coupling interactions between equally  pumped polariton condensates assuming an exponential  profile of the individual polariton densities. For unequally populated polariton condensates the coupling strength is derived from a series expansion  with respect to their outflow wave-vector difference. We analyse the limit under which two condensates remain coherently coupled. We show that phase configuration maximising the polariton occupancy across a dyad corresponds to the minimisation of the sum of the symmetric Heisenberg exchange and asymmetric Dzyloshinskii-Moriya interactions \citep{dm}. We show that using the non-equal pumping in a polariton dyad to control the  coupling between the condensates allows accessing a wider class of optimization problems than equal pumping.


\section{Symmetric Heisenberg and asymmetric Dzyaloshinskii-Moriya interactions}

In the following, we investigate the case of two unequally pumped exciton reservoirs with Gaussian pumping profiles and explore the limit of coherent coupling across the dyad.  For two spatially separated condensates, we approximate the wave-function of the system as the sum of the two wavefunctions of the individually created condensates:
\begin{equation}
	\psi({\bf r}) \approx \Psi_1(|{\bf r} - {\bf r_1}|) + \Psi_2(|{\bf r} - {\bf r_2}|),
\end{equation}
where the wavefunction of a condensate located at ${\bf r}  = {\bf r_i}$ can be approximated \cite{BerloffNJP2017} by
\begin{equation}
	\Psi_i(|{\bf r - r_i}|) \approx  \sqrt{\rho_i(|{\bf r - r_i}|)} \exp[{\rm i} k_{ci} |{\bf r - r_i}| + i \theta_i],
\end{equation}
where $\theta_i$ is the space independent part of the phase,  $k_{ci}$ is the maximum wave-vector $k(\textbf{r})$ that polaritons reach within their lifetime by converting their potential to kinetic energy  \citep{QuantumFluidsOfLight2013}, and $\rho_i(|{\bf r- r}_i|)$ is the density of the isolated condensate created by a single pumping source centered at ${\bf r}_i$; for an approximate expression for $\rho_i$ as a function of the system parameters  see  \cite{BerloffNJP2017}. 

The total number of polaritons across the dyad is given by $\mathcal{N} = \int |\psi({\bf r})|^2 d{\bf r}$, where integration is over the entire area of the microcavity and 
\begin{eqnarray}
	\mathcal{N} &\approx &  \int |  \Psi_1(|{\bf r - r_1}|)+\Psi_2(|{\bf r - r_2}|)|^2 d{\bf r} = \nonumber \\
	&=& \int d{\bf r} \left[ |\Psi_1(|{\bf r - r_1}|)|^2+  |\Psi_2(|{\bf r - r_2}|)|^2 \right] + \nonumber \\
	&+& \int d{\bf r}  \left[ \Psi_1(|{\bf r - r_1}|) \Psi_2^*(|{\bf r - r_2}|) + c.c. \right] = \nonumber  \\
	&=& \mathcal{N}_1 + \mathcal{N}_2 + J \cos \Delta \theta + D \sin \Delta \theta,
	\label{N_samespins}
\end{eqnarray}
where $\Delta \theta= \theta_1 - \theta_2$ is the phase difference between two polariton condensates, $\mathcal{N}_i=\int |\Psi_i(|{\bf r-r}_i|)|^2\, d{\bf r}$  is the number of polaritons of an individual condensate indexed by $i$, and the interaction strengths $J$ and $D$ are expressed as
\begin{eqnarray}
J &=& 2 \int \sqrt{\rho_1 (|{\bf r} - {\bf r_1}|)} \sqrt{\rho_2 (|{\bf r} - {\bf r_2}|)}  \nonumber \\
&\times& \cos [k_{c1} |{\bf r} - {\bf r_1}| - k_{c2} |{\bf r} - {\bf r_2}|] d {\bf r},
\label{Initial_J}\\
D &=& 2 \int \sqrt{\rho_1 (|{\bf r} - {\bf r_1}|)} \sqrt{\rho_2 (|{\bf r} - {\bf r_2}|)}  \nonumber \\
&\times& \sin[k_{c1} |{\bf r} - {\bf r_1}| - k_{c2} |{\bf r} - {\bf r_2}|] d {\bf r}.
\label{Initial_D}
\end{eqnarray}
From all the possible phase differences, $\Delta \theta=[0,2\pi)$, the one that maximises the number of particles in Eq. (\ref{N_samespins}), will condense first as was established in \cite{OGFR} for the equally pumped polariton dyad. Equivalently, in the generic  case of a polariton dyad with unequal populations, the system will reach threshold at the phase difference configuration $\Delta\theta$ that minimises
\begin{equation}
H_T=-(J \cos \Delta\theta + D \sin \Delta\theta),
\label{Total_H}
\end{equation}
where $H_T$ is the sum of the symmetric Heisenberg exchange and the asymmetric Dzyaloshinskii-Moriya (DM) interactions \cite{dm}, that are usually studied in the context of a contribution to the total magnetic exchange interaction between two neighboring magnetic spins \citep{DMinteractions1983}. 

We can draw here an analogy with  the superexchange interactions between two neighboring sites established in BECs of ultracold atoms loaded into optical lattices \cite{trotzky08} that allowed to carry out the analog simulation of frustrated classical magnetism in triangular optical lattices \cite{OpticalLatticesScience2011}. In addition, the spin-orbit coupling in optical lattices can give rise to the DM spin interactions \cite{so}, therefore, allowing the simulation of  spiral
order and multiferroic effects. In the case of polariton condensates, therefore,  the unequal pumping allows to simulate these effects as well.

\section{Analytical expressions for the coupling strengths}

In the following, we obtain analytical expressions of the coupling strengths $J$ and $D$ by positioning the condensates at $\textbf{r}_1 = (-d/2,0)$ and $\textbf{r}_2 = (d/2,0)$, where $d=|{\bf r_1-r_2}|$ is the separation distance, and transforming into elliptic coordinates $(\mu, \nu)$ with
\begin{eqnarray}
	x &=& \frac{d}{2} \cosh \mu \cos \nu, \\
	y &=& \frac{d}{2} \sinh \mu \sin \nu, \\
	d^2 {\bf r} &=& \frac{d^2}{4} (\sinh^2 \mu + \sin^2 \nu) d\mu d \nu,
\end{eqnarray}
where 
$\mu$  is a nonnegative real number and 
$\nu \in [0, 2\pi)$, so that  the expressions for the absolute values simplify  to
\begin{eqnarray}
	|{\bf r} - {\bf r_1}| &=& \frac{d}{2} (\cosh \mu + \cos \nu),
	\label{d1} \\
	|{\bf r} - {\bf r_2}| &=& \frac{d}{2} (\cosh \mu - \cos \nu).
	\label{d2}
\end{eqnarray}
Assuming an exponential decay of the amplitude for an individual condensate $\sqrt{\rho_i (|{\bf r} - {\bf r_i}|)} = A_i \exp(-\beta |{\bf r} - {\bf r_i}|)$, where $A_i$ and $\beta$ correlate with  the shape of the pumping profile \cite{BerloffNJP2017}, and substituting Eqs. (\ref{d1})-(\ref{d2}) we obtain
\begin{eqnarray}
	J &=& \frac{1}{2} A_1 A_2 d^2 \int_0^{\infty} e^{- \beta d \cosh \mu} \int_0^{2\pi} (\sinh^2 \mu + \sin^2 \nu)   \nonumber \\
	&\times &   \cos (\delta k_c^- d \cosh \mu - \delta k_c^+ d \cos \nu) d\nu d \mu, \\
	D &=& \frac{1}{2} A_1 A_2 d^2 \int_0^{\infty} e^{- \beta d \cosh \mu} \int_0^{2\pi} (\sinh^2 \mu + \sin^2 \nu)  \nonumber \\
	&\times &   \sin (\delta k_c^- d \cosh \mu - \delta k_c^+ d \cos \nu) d\nu d \mu,
\end{eqnarray}
where we denoted $\delta k_c^{\pm} = (k_{c1} \pm k_{c2}) /2$. Integrating firstly over $\nu$ and then expanding the  integrand for small $\delta k_c^-$ up to the third order, we obtain analytical expressions for the coupling strengths of the two unequally pumped condensates in terms of  the Bessel  functions ($J_n$) and the modified Bessel functions of the second kind ($K_n$):
\begin{eqnarray}
	J &=& \pi A_1 A_2 d \bigg[ \frac{1}{\beta} J_0(\delta k_c^+ d) K_1(\beta d)   \nonumber \\
	&+& \frac{1}{\delta k_c^+} J_1(\delta k_c^+ d) K_0(\beta d) - (\delta k_c^-)^2  F  \bigg],
	\label{J_analytical} \\
	D &=& \pi A_1 A_2 d^2  \delta k_c^- \bigg[ \frac{1}{\beta} J_0(\delta k_c^+ d) K_2(\beta d)  \nonumber \\
	&+& \frac{1}{\delta k_c^+ } J_1(\delta k_c^+ d) K_1 (\beta d) - (\delta k_c^-)^2 G \bigg],
	\label{D_analytical}
\end{eqnarray}
where
\begin{eqnarray}
F &=& \frac{d}{2 \beta^2} J_0(\delta k_c^+ d) \big\{ \beta d K_1(\beta d) + 3K_2(\beta d)   \big\} +  \nonumber \\
	&+& \frac{d}{2 \beta} \frac{J_1(\delta k_c^+ d)}{\delta k_c^+} \big\{  ( \beta d K_0(\beta d) + K_1(\beta d) \big\}, \\
G &=& \frac{d}{6 \beta^2} J_0(\delta k_c^+ d) \big\{ \beta d K_2(\beta d) + 3K_3(\beta d)   \big\} +  \nonumber \\
	&+& \frac{d}{6 \beta} \frac{J_1(\delta k_c^+ d)}{\delta k_c^+} \big\{  ( \beta d K_1(\beta d) + K_2(\beta d)  \big\},
\end{eqnarray}
We note here that the integrals can be analytically calculated up to any desired precision of $(\delta k_c^-)^n$. An exact analytical expression for two equally pumped polariton condensates with $A_1 = A_2 = A$ and $k_{c1} = k_{c2} = k_c$ ($D=0$) reads
\begin{equation}
	J = \pi A^2 d  \bigg[ \frac{1}{\beta} J_0 (k_c d) K_1 (\beta d) +  \frac{1}{k_c} J_1 (k_c d) K_0 (\beta d)\bigg].
\end{equation}
If the pumping width is large ($\beta$ is small) the sign of the interactions is determined by $J_0(k_c d)$ as was found in \cite{OGFR}.

\begin{figure}[t!]
\centering
  \includegraphics[width=8.6cm]{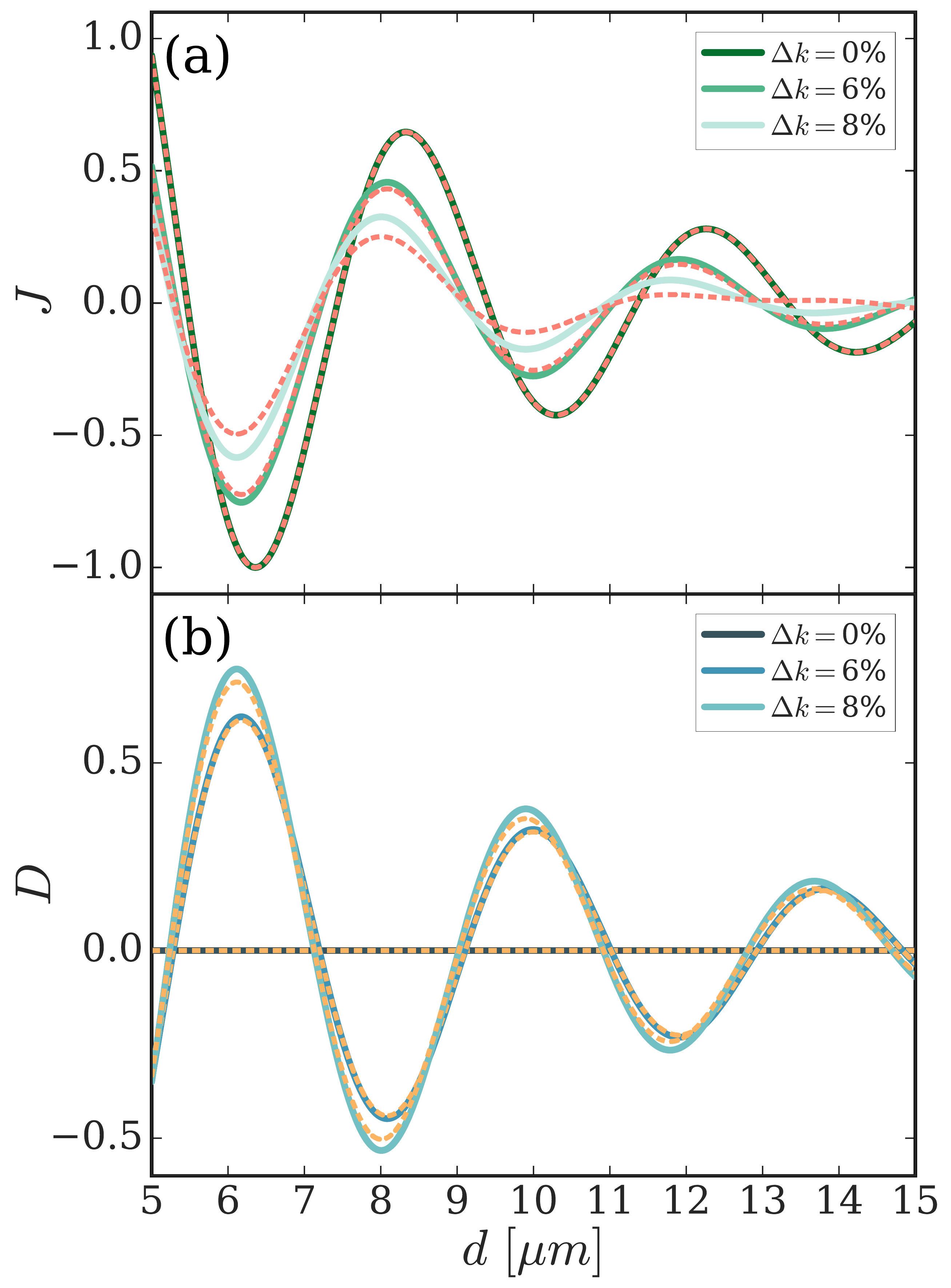}
\caption{The strengths of (a) the symmetric Heisenberg exchange $J$ and (b) the asymmetric Dzyaloshinskii-Moriya  interactions  $D$ as functions of the  separation distance $d$. The solid (dashed) lines show the normalised  coupling strengths found   numerically (analytically)  from Eqs. (\ref{Initial_J})-(\ref{Initial_D}) (Eqs. (\ref{J_analytical})-(\ref{D_analytical})). Colours correspond to the different population imbalances characterized  by the percentage differences of the condensates'  wavevectors, $\Delta k=0\%$, $6\%$ and $8\%$, where $\Delta k= 100\% \cdot (k_{c2}-k_{c1})/k_{c1}$. The parameters are $k_{c1} = 1.6\mu m^{-1}$, $\beta = 0.2\mu m^{-1}$.}
 \label{Figure1}
\end{figure}

Figure \ref{Figure1} shows the analytically and numerically calculated $J$ and $D$ as functions of the distance separating two condensates for small differences between the outflow wavevectors. The agreement improves  even further when higher orders of $\delta k_c^-$ in (\ref{J_analytical}-\ref{D_analytical}) are taken into account.  We note that  a  discrepancy  between the polariton wavevectors $k_{c1}$ and $k_{c2}$ may lead to  significant non-zero values of the coupling strength $D$ that may even exceed values of  $J$. In particular, for the range of experimental parameters  it is possible to obtain a continuous phase transition between anti-ferromagnetic coupling for equal pumping ($\Delta\theta= \pi$, $J<0$ and $D = 0$) and ferromagnetic coupling  for unequal pumping ($\Delta\theta = 0$, $J \approx 0$ and $D>0$).
\begin{figure}[t!]
\centering
 \includegraphics[width=8.6cm]{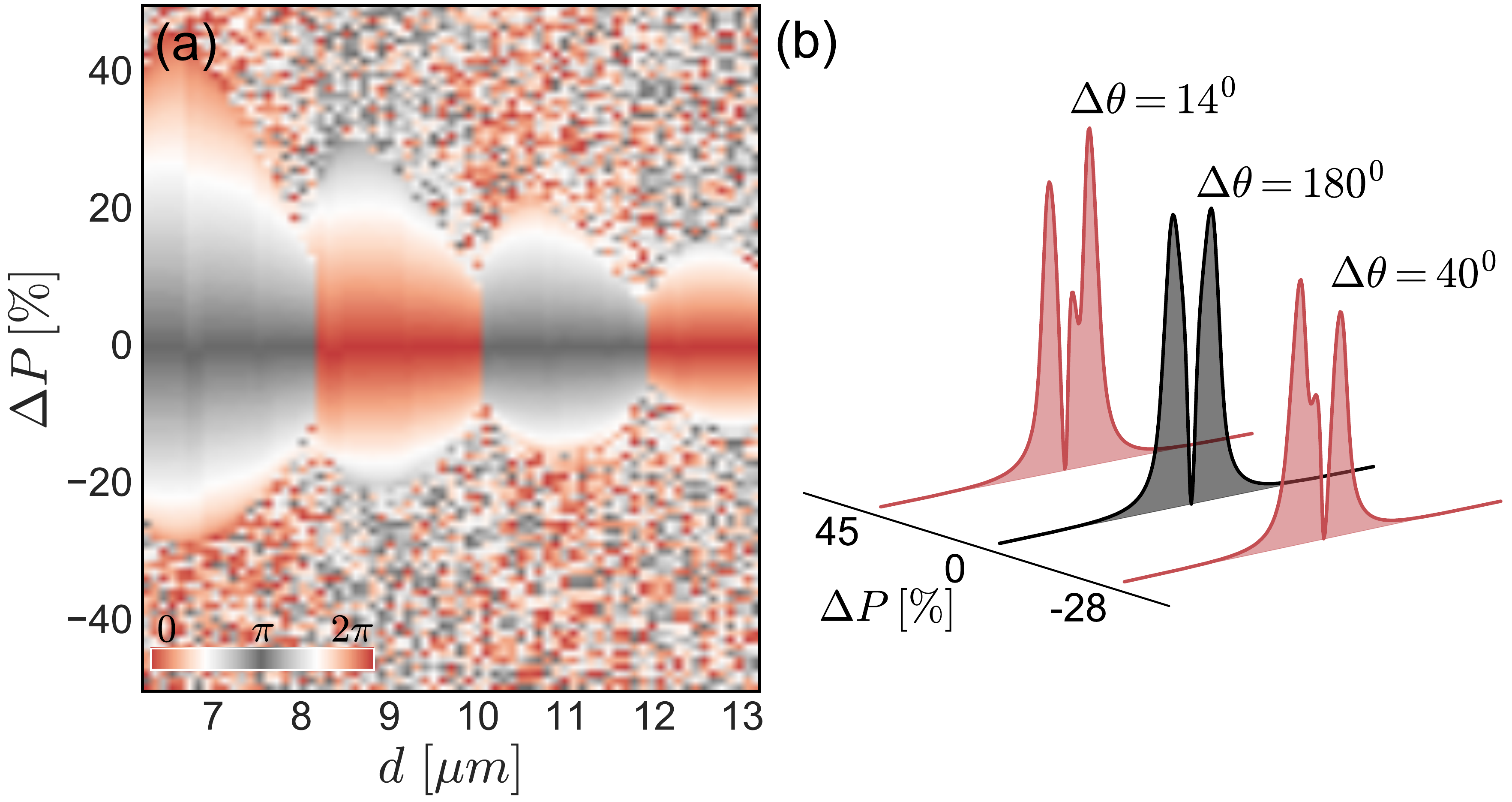}
\caption{(a) Contour plot of the phase difference $\Delta\theta = \theta_1-\theta_2$ in the polariton dyad as a function of the pumping imbalance $\Delta P = 100\% \cdot (P_2 - P_1) / P_1 $ and the separation distance $d$. The pumping corresponding to $P_1$ is the same used in our experimental work \cite{BerloffNatMat2017} and is kept fixed. The phase difference is measured either for the steady state or at a fixed moment of time for time-dependent oscillations that characterize the loss of phase locking between the condensates.  Fragmentation of the contour plot, therefore, represents the region with the loss of the phase locking (coherence) between the condensates.    (b) Density cross-sections of the polariton dyad  along the line connecting the condensate centers positioned at the distance $d = 6.4 \mu m$ apart and pumped with $\Delta P = \{-28,0,45\}\%$.  }
 \label{Figure2_phasemap}
\end{figure}

\section{Loss of coherence in a polariton dyad}
To determine the levels of the pumping imbalance and the distances for which  polariton condensates in a polariton dyad remain coherently coupled  we use the complex Ginzburg-Landau equation (cGLE) with a saturable nonlinearity \citep{Wouters, Berloff} written for a polariton wavefunction $\psi$ in 2D:
\begin{eqnarray}	
	i \hbar  \frac{\partial \psi}{\partial t} &=& - \frac{\hbar^2}{2m}  \left(1 - i \eta_d {\cal R} \right) \nabla^2\psi + U_0 |\psi|^2 \psi +  \nonumber \\
	  &+& \hbar g_R {\cal R} \psi + \frac{i\hbar}{2} \biggl(R_R {\cal R} - \gamma_C \biggr) \psi, \label{Initial GL equation}\\
	  \frac{\partial \cal R}{\partial t} &=&  - \left( \gamma_R + R_R |\psi|^2 \right) {\cal R} + P({\bf r}),
	\label{Initial Reservoir equation}
\end{eqnarray}
where $m$ is the effective polariton mass, $U_0$ and $g_R$ are the polariton-polariton and exciton-polariton interaction strengths respectively, $R_R$ is the transfer rate from the exciton reservoir to the polariton condensate, $\gamma_C$ is the polariton radiative decay rate, $\gamma_R$ is the exciton reservoir non-radiative decay rate, ${\cal R}$ is the density of the exciton reservoir, $\eta_d$ is the energy relaxation term, and $P$ is the pumping profile replenishing the exciton reservoir.  Previously, we calibrated the parameters in  Eq. (\ref{Initial Reservoir equation}) using an extensive set of  experimental data \cite{BerloffNatMat2017}, so that we can take the results of the numerical simulations as the experimentally representative results.  For the range of pumping parameters where there is a mutual coherence across the polariton dyad, the condensates share the same global chemical potential $\mu_{coh}$, which can be found from Eq. (\ref{Initial GL equation})  by substituting $\psi \rightarrow \psi \exp(- i \mu_{coh} t / \hbar)$.  For each separation distance $d$ we numerically integrate  Eq. (\ref{Initial Reservoir equation}) starting with many random phase difference configurations and choosing the one that maximizes the total number of particles as this  corresponds to the state that will lase first \cite{OGFR}.
 We depict the resulting phase differences in Fig.~\ref{Figure2_phasemap}(a). Here the phase difference $\Delta\theta$ in the polariton dyad is shown in a color scale as a function of the pumping imbalance  $\Delta P $ and the separation distance $d$.  If two condensate in a dyad are coherently coupled they share the same chemical potential and achieve a steady state. Above some critical pumping imbalance (which is different for different separation distances)  the coherence is lost, each condensate has its own chemical potential and the phase difference oscillates in time. This region is shown as fragmented on Fig.  \ref{Figure2_phasemap}(a).    
Figure \ref{Figure2_phasemap}(a) demonstrates that the whole range of phase differences from $0$ to $\pi$ is  achievable without losing coherence in a polariton dyad by  adjusting the intensity of one of the pumps and the distance  between the condensates. Figure \ref{Figure2_phasemap}(b) depicts the density cross-sections of the polariton dyad  along the line connecting the condensate centers. The change in the pumping intensity of one of the spots changes the phase locking from $\pi$ phase difference configuration to almost ferromagnetic configurations. 

\begin{figure}
\centering
 \includegraphics[width=8.6cm]{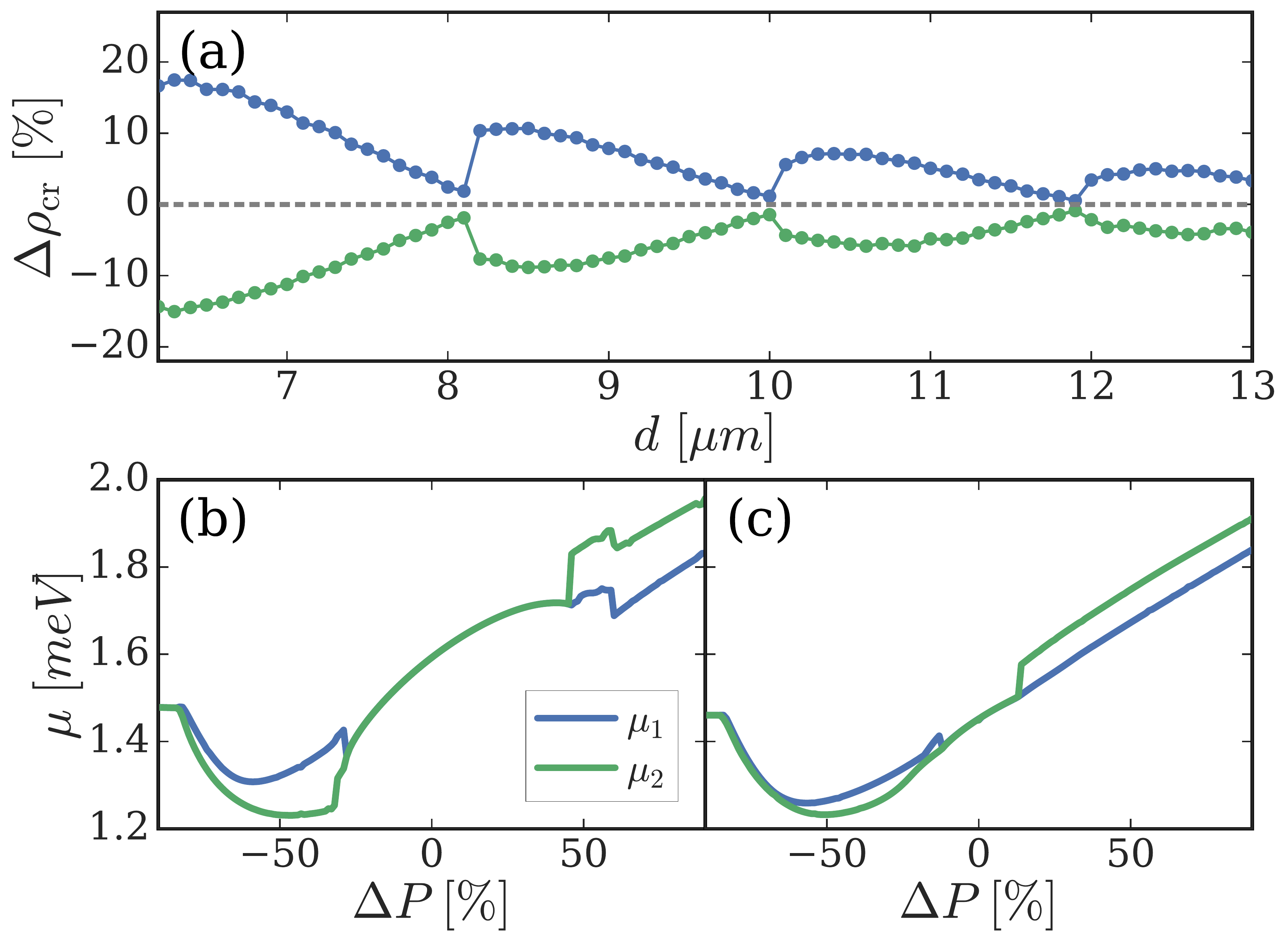}
\caption{ (a) The critical population imbalance for the loss of coherence in polariton dyad $\Delta \rho_{cr}=100\% \cdot (|\psi({\bf r}_1)|^2 - |\psi({\bf r}_2)|^2)/|\psi({\bf r}_2)|^2$ for mutually coherent condensates in a polariton dyad as a function of the separation distance $d$.  The chemical potentials of polariton condensates in  a polariton dyad as  functions of $\Delta P = 100\% \cdot (P_2 - P_1) / P_1 $ for the distances (b) $d = 6.4 \mu m$ and (c) $d = 13 \mu m$.  Condensates share the same chemical potential  where two lines join together in a finite region around $\Delta P =0$. Beyond this region the condensates become decoupled and acquire different $\mu$.}
 \label{Figure4_wavevectors}
\end{figure}

The critical values of the population imbalance  in the polariton dyad, under which the condensates will remain coherent, are provided in Fig.~\ref{Figure4_wavevectors}(a) where $\Delta \rho_{cr} $ between the maxima of the polariton densities  at the threshold for phase decoupling is shown  as a function of  the separation distance $d$. We also calculated the chemical potentials of the condensates   in a dyad for the two particular separation distances  $d = 6.4 \mu m$  and  $d = 13 \mu m$ shown in  Fig.~\ref{Figure4_wavevectors}(b). The condensates are coherently coupled  for small pumping imbalances  in a finite region around  $\Delta P=0$. Beyond this region  the  mutual coherence is lost  and each condensate acquires its own chemical potential. The appearance of such splitting manifests itself via time-dependent phase difference oscillations. 

\section{Analog Hamiltonian simulator  and quadratic minimization problems}
If $n$ condensates are pumped with the same intensity they achieve the global minimization of the  classical $XY$ Hamiltonian  $H_{XY}=-\sum_{i=1}^n J_{ij} \cos (\theta_i-\theta_j)$ \citep{BerloffNatMat2017}. By writing $z_j=\cos\theta_i + {\rm i} \sin\theta_j$ we convert the minimization of the XY Hamiltonian into the constant modulus constrained quadratic minimization problem 
\begin{equation}
\min  {\bf z}^H {\bf Q} {\bf z}, \quad {\rm subject} \quad {\rm to} \quad  |z_j| = 1,
\label{qm}
\end{equation}
 where ${\bf z}=\{z_j\}$ is a complex $n \times 1$ vector, and ${\bf Q}$ is a real symmetric $n \times n$ matrix with the elements ${\bf Q}=\{-J_{ij}\}, i\ne j$, and $Q_{ii}=0$.

  If $n$ condensates are pumped with different intensities the Hamiltonian given by Eq. (\ref{Total_H}) can be generalized to the case of $n$ condensates centered  at ${\bf r}_i$, $i=1,...,n$ to
  \begin{equation}
  H_T= -\sum_{i=1}^n J_{ij} \cos (\theta_i-\theta_j) - \sum_{i=1}^n D_{ij} \sin (\theta_i-\theta_j).
  \end{equation}
  This represents the constant modulus complex quadratic optimisation problem 
  \begin{equation}
\min  {\bf z}^H {\bf G} {\bf z}, \quad {\rm subject} \quad {\rm to} \quad  |z_j| = 1,
\label{qm2}
\end{equation}
with ${\bf G}
  $ being a Hermitian matrix with elements ${\bf G}=-\{J_{ij} \pm {\rm i}D_{ij}\}, i\lessgtr j$, 
  and $G_{ii}=0$. 

Such a Hermitian quadratic optimization model appears  in application to signal processing \cite{luo} and is known as phase retrieval problem  which arises in X-ray tomography of biomedical imaging \cite{xray},
astronomical imaging \cite{astro}, diffraction imaging \cite{diff},  optics \cite{optics}, microscopy \cite{micro}, and many other applications \cite{otherPR}. Phase retrieval is the fundamental problem of recovering a general signal (or image) from the
magnitude of its Fourier transform. The signal detectors can often  record only the squared modulus of the Fresnel or Fraunhofer diffraction pattern
of the radiation that is scattered from an object. So the information about the phase of the optical
wave reaching the detector is lost  making the information about the scattered object or the optical field to be incomplete. To extract the information about the phases usually requires  a priori information about the signal, such as positivity, real-valuedness, atomicity, support
constraints,  and so on \cite{assum}, since the large computational complexity of the phase retrieval problems limits the application of direct methods to small-scale problems.  Indeed, it was proved in \cite{zhang} that the optimization problem (\ref{qm2}) is strongly NP-hard in general even for a semi-positive Hermitian matrix ${\bf G}$. For this problem even finding an approximate solution  using semidefinite programming and  the randomization-rounding
procedure guarantees a worst-case performance ratio of $\pi/4$ \cite{zhang}.  Polariton condensates arranged in a graph and interacting via symmetric Heisenberg and non-symmetric DM coupling can therefore be used as an analog solver for such hard problems. 

\section{Conclusions}

In conclusion, we study the coherence and phase locking  between  spatially separated polariton condensates that are pumped  nonresonantly at various intensity. Depending on the separation
distance between condensates, the pumping difference and the flow velocities of 
 polaritons two condensates in a polariton dyad may  either synchronize with
a phase difference in $[0, \pi)$ or become decoherent with the phase difference undergoing the time-dependent oscillations. This is in a contrast with the equally pumped polariton dyad where the condensates are always coherent with phase locking with $0$ or  $\pi$ phase difference.  We derive the  analytic  expressions for the coupling strengths. For equally pumped condensates we account for the finite width of the pumping configuration to make an improvement of the previously found estimates \cite{OGFR}. In the case of non-equal pumping the analytic expressions for the coupling strengths are derived as a series expansion in terms of the outflow velocity differences.  Using the numerical integration of the governing mean field equations we calculated the phase diagram for the phase difference in a polariton dyad indicating the region where the coherence is lost.   Our theoretical predictions could potentially broaden the range of optimisation problems that can be addressed with polariton optimizers. 

\section{Aknowledgements}
The authors acknowledge the support of the Skoltech NGP Program (Skoltech-MIT joint project).

\end{document}